\documentstyle[12pt,aasms4]{article}
\makeatletter

\@ifundefined{chapter}{\def\thebibliography#1{\section*{References\@mkboth
  {REFERENCES}{REFERENCES}}\list
  {\relax}{\setlength{\labelsep}{0em}
        \setlength{\itemindent}{-\bibhang}
        \setlength{\itemsep}{0pt}
        \setlength{\parsep}{0pt}
        \setlength{\leftmargin}{\bibhang}}
    \def\newblock{\hskip .11em plus .33em minus .07em}
    \sloppy\clubpenalty4000\widowpenalty4000
    \sfcode`\.=1000\relax}}%
{\def\thebibliography#1{\chapter*{Bibliography\@mkboth
  {BIBLIOGRAPHY}{BIBLIOGRAPHY}}\list
  {\relax}{\setlength{\labelsep}{0em}
        \setlength{\itemindent}{-\bibhang}
        \setlength{\itemsep}{0pt}
        \setlength{\parsep}{0pt}
        \setlength{\leftmargin}{\bibhang}}
    \def\newblock{\hskip .11em plus .33em minus .07em}
    \sloppy\clubpenalty4000\widowpenalty4000
    \sfcode`\.=1000\relax}}

\newlength{\bibhang}
\let\@internalcite\cite
\def\cite{\let\@citeleft(\let\@citeright)%
    \@ifstar{\citeyear}{\citefull}}
\def\acite{\let\@citeleft\relax\let\@citeright\relax%
    \@ifstar{\citeyear}{\acitefull}}
\def\citenp{\let\@citeleft\relax\let\@citeright\relax
    \@ifstar{\citeyear}{\citefull}}
\def\citefull{\def\astroncite##1##2{##1~##2}\@internalcite}
\def\citeyear{\def\astroncite##1##2{##2}\@internalcite}
\def\acitefull{\def\astroncite##1##2{##1~(##2)}\@internalcite}

\def\@citex[#1]#2{\if@filesw\immediate\write\@auxout{\string\citation{#2}}\fi
  \def\@citea{}\@cite{\@for\@citeb:=#2\do
    {\@citea\def\@citea{; }\@ifundefined
       {b@\@citeb}{{\bf ?}\@warning
       {Citation `\@citeb' on page \thepage \space undefined}}%
{\csname b@\@citeb\endcsname}}}{#1}}

\def\@cite#1#2{\@citeleft#1\if@tempswa , #2\fi\@citeright}
\def\@biblabel#1{}

\makeatother

\newcommand{\PSbox}[3]{\mbox{\rule{0in}{#3}\includegraphics{#1}\hspace{#2}}}
\newcommand{\FigNum}[1]{\unitlength 1pt \begin{picture}(55,10)(-400,35) 
                        \put(0,0){Figure #1}
                        \end{picture}}
 
\newcommand{\zsun}{$Z_\odot$} 
\newcommand{\persec}{\mbox{$\second^{-1}$}}
\newcommand{\percm}{\mbox{$\cm^{-2}$}}
\newcommand{\ppm}{\mbox{$\pm$}}
\newcommand{\cgsflux}{\erg\percm\persec}
\newcommand{\cgslum}{\erg\persec}

\newcommand\approxlt{\mbox{$^{<}\hspace{-0.24cm}_{\sim}$}}
\def\etal{{et~al.}}

\newcommand{\nh}{\mbox{$N_{\rm H}$}}
\newcommand{\nhtt}{\mbox{$N_{\rm H, 22}$}}
\newcommand{\ud}[2]{\mbox{$^{+ #1}_{- #2}$}}
\newcommand{\ee}[1]{\mbox{$10^{#1}$}}
\newcommand{\tee}[1]{\mbox{$\times 10^{#1}$}}

\newcommand\lx{$L_{X}$}

\newcommand{\perval}[2]{{#1\mbox{$^{#2}$}}} 

\def\chisqrnu{\mbox{$\chi^2_\nu$}}
\def\x1608{{4U~1608$-$522}}

\def\cenx4{{Cen~X$-$4}}
\def\aql{{Aql~X$-$1}}
\def\saxj1808{{SAX J1808.4$-$3658}}

\newcommand{\cm}{\mbox{$\rm\,cm$}}

\newcommand{\second}{\mbox{$\rm\,s$}}

\newcommand{\erg}{\mbox{$\rm\,erg$}}

\newcommand{\kteff}{\mbox{$kT_{\rm eff}$}}
\newcommand{\kteffinfty}{$kT_{\rm eff}^\infty$}

\newcommand{\rinfty}{\mbox{$R_{\infty}$}}

\newcommand{\chandra}{{\em Chandra\/}}

\newcommand{\rosat}{{\em ROSAT\/}}

\def\ngc{NGC~5139}
\def\llxs{{LLXS}}
\def\source3{CXOU~132619.7-472910.8}
\def\ktbremss{\mbox{$kT_{\rm bremss.}$}}
\newcommand{\LH}{2}
\newcommand{\SingleSpace}{
  \renewcommand{\LH}{0.90}
  \def\baselinestretch{\LH}
  \tiny
  \normalsize
}
\begin{document}

\title{Identification of A Transient Neutron Star in
Quiescence in the Globular Cluster NGC 5139}

\author{Robert E. Rutledge\altaffilmark{1}, 
Lars Bildsten\altaffilmark{2}, Edward F. Brown\altaffilmark{3}, 
George G. Pavlov\altaffilmark{4}, 
\\ and Vyacheslav  E. Zavlin\altaffilmark{5}}
\altaffiltext{1}{
Space Radiation Laboratory, California Institute of Technology, MS 220-47, Pasadena, CA 91125;
rutledge@srl.caltech.edu}
\altaffiltext{2}{
Institute for Theoretical Physics and Department of Physics, Kohn Hall, University of 
California, Santa Barbara, CA 93106; bildsten@itp.ucsb.edu}
\altaffiltext{3}{
Enrico Fermi Institute, 
University of Chicago, 
5640 South Ellis Ave, Chicago, IL  60637; 
brown@flash.uchicago.edu}
\altaffiltext{4}{
The Pennsylvania State University, 525 Davey Lab, University Park, PA
16802; pavlov@astro.psu.edu}
\altaffiltext{5}{ 
Max-Planck-Institut f\"ur Extraterrestrische Physik, D-85740 Garching,
Germany; zavlin@xray.mpe.mpg.de}

\begin{abstract}

 Using the \chandra/ACIS-I detector, we have identified an X-ray
source (\source3) in the globular cluster \ngc\ with a thermal
spectrum identical to that observed from transiently accreting neutron
stars in quiescence. The absence of intensity variability on
timescales as short as 4 seconds ($<$ 25\% rms variability) and as
long as 5 years ($<$50\% variability) supports the identification of
this source as a neutron star, most likely maintained at a high
effective temperature ($\approx 10^6 \ {\rm K}$) by transient
accretion from a binary companion. The ability to spectrally identify
quiescent neutron stars in globular clusters (where the distance and
interstellar column densities are known) opens up new opportunities
for precision neutron star radius measurements.

\end{abstract}

\section{Introduction}
\label{sec:intro}

The large number of low luminosity X-ray sources (\llxs ; \lx
\approxlt \ee{34} \cgslum) in globular clusters (GCs) were initially
explained as accreting white dwarfs \cite{hertz83}, although some
fraction may be active RS~CVn binaries \cite{bailyn90}. Some of these
systems may also be transient neutron stars in quiescence (qNSs;
\citenp{verbunt84}).  The observations of transient neutron stars in
outburst in globular cluster NGC 6440
\cite{forman76b,intzand99,verbunt00b} and in Liller 1 \cite{lewin76},
supports the suggestion that such a population exists in globular
clusters.  We report here on our spectral search for such an object
within the globular cluster \ngc ($\omega$ Cen).

Much observational and theoretical work has been carried out on
accreting transients in the field, most of which are black holes. The
quiescent emission from neutron stars has been well observed,
resulting in numerous models for the source of the quiescent emission,
ranging from accretion in quiescence to thermal emission from the NS
surface (see \citenp{bildstenrutledge00} for an overview). Brown,
Bildsten \& Rutledge \cite*[BBR98 hereafter]{brown98} showed that the
neutron star (NS) core is heated by reactions deep in the NS crust
during the repeated accretion outbursts common in transients (for
reviews of transient neutron stars, see \citenp{chen97,campana98b}).
The core is heated to a steady-state temperature on a timescale of
$\sim$\ee{4} yr (see also \citenp{colpi00}), providing an emergent
thermal luminosity (BBR98):
\begin{equation}
\label{eq:brown}
L_q = 8.7\times10^{33} \left(\frac{\langle \dot{M} \rangle}{10^{-10}
M_\odot {\rm yr}^{-1}}\right) \frac{Q}{1.45 {\rm  MeV}/m_p} \; \; {\rm ergs \; s}^{-1}
\end{equation}
where $Q$ is the nuclear energy release in the crust \cite{haensel90}. 

The soft X-ray spectral component of quiescent neutron stars is
expected to be an emergent H atmosphere spectrum (BBR98; for
calculations of these spectra, see \citenp{rajagopal96,zavlin96}).
Spectral fits assuming a H atmosphere spectrum yield emission areas
consistent with a NS (Rutledge \etal\ 1999, 2000, 2001a,
2001b). \nocite{rutledge99,rutledge00,rutledge01,rutledge01b} In
addition, a hard spectral component often dominates the spectrum above
2 keV in the two field qNSs \cenx4\
\cite{asai96b,asai98,campana00,rutledge01} and \aql\
\cite{campana98a,rutledge01b}, the origin of which is still not clear
\cite{campana98b}.  

  We have examined archived \chandra/ACIS-I observations of the
globular cluster NGC 5139 to spectrally identify qNSs. One X-ray
source stands out as having a spectrum consistent with that from a
pure hydrogen atmosphere of a 10 km neutron star.  This source is also
spectrally inconsistent with millisecond radio pulsars (MSPs),
cataclysmic variables (CV), or RS~CVn binaries.  An interpretation of
this source as a narrow line Seyfert 1 (NLSy1) is not favored, as it
requires an atypical X-ray spectrum for this class, and requires
accepting as a coincidence that the intensity should be the same as from
a 10 km neutron star at the distance of \ngc.

In \S~2 we describe the observation and spectral analysis of the point
sources in the field of \ngc, comparing the X-ray spectra to models
appropriate to MSPs, CVs, RS~CVns and NLSy1s.  We examine
previous observations of the identified qNS in \S~3, and conclude in
\S~4 with a discussion. 

\section{Observation and Point Sources}

The \chandra/ACIS-I detectors observed \ngc\ in imaging mode for two
continuous periods (2000 Jan 24 02:15-09:46  and Jan 25
04:33-17:24 TT) for a total exposure of $\approx$ 68.6 ksec.  We combined
the two observations and analyzed them as one, using CIAO
v2.1\footnote{http://asc.harvard.edu/ciao/} and XSPEC v11
\cite{xspec}. We searched for point sources using {\em celldetect},
using only the 0.1-2.5 keV energy range, excluding regions $<$16
pixels from the detector edges, and keeping only sources with
S/N$>$5.0.  We find 40 X-ray point sources over the four ACIS-I chips
and chip S2 of ACIS-S, which are numbered in order of decreasing S/N
in Table~\ref{tab:sources}. X-ray point-source counts from
extra-galactic deep surveys \cite{hasinger98,giacconi01} imply about
40 background sources total in the 345 sq arcmin region, down to the
approximate limit of our sample, 2.6\tee{-15} \cgsflux (0.5-2 keV).
Hence, only a fraction of the detected sources are likely to be
members of the GC.

  We excluded the sources \#12, 18, 28 and 30 from spectral analysis
due to their close proximity ($<$33 pixels) to the detector edge.  We
also excluded sources \#1, 9, 16, and 36, as these were $>$12\arcmin\
off-axis, with large point spread functions offset in position from
{\em celldetect} position by up to 10\arcsec perhaps due in part to
the low S/N. For all other point sources, we extracted source counts
within a radius $r=2.1 + 0.073({R}/{\rm 1 \; arcmin})^{2.32}\; {\rm
arcsec}$ where $R$ is the distance from the ACIS-I aimpoint; 
this is slightly larger than the 90\% enclosed energy radius at 6
keV.  Background counts were extracted from an annulus centered on
the source position, with an inner radius 2 pixels larger than the
source region, and an outer radius which was the larger of a factor of
3 times the inner-radius, or 50 pixels, excluding overlapping source
regions and off-chip regions. We used the script {\em psextract} to
extract the source counts and background, to generate pulse-invariant
(PI) spectra for analysis.  We binned the PI spectra into four energy
bins: 0.5-1 keV, 1-2 keV, 2-5 keV, and 5-10 keV.

 Even though most of these sources are bound to be active galaxies, 
we begin by comparing these X-ray spectra with spectral models for qNSs,
CVs, MSPs and RS CVn's. All spectral models
include galactic absorption at the optically implied value of
\nhtt=0.09 \cite{djorgovski93}, which is comparable to the 21 cm line
value \cite{dickey90}.  We assume a fiducial distance to \ngc\ of 5
kpc, between the extremes of 4.9 and 5.3 kpc which are typically
quoted \cite{djorgovski93,harris96}.

  We begin by assuming a H atmosphere spectrum and holding \rinfty=13
km (\rinfty=$R/\sqrt{1 - 2M/(Rc^2)}$, where R is the proper radius,
$M$ is the NS mass, and $c$ is the speed of light). We then spectrally
fit the PI spectra of all X-ray sources.  Of the 32 spectrally
analyzed sources, two (which we label source \#3=\source3\ and \#31)
were acceptably fit (prob(\chisqrnu)$>$0.01; \chisqrnu is the reduced
chi-square statistic; \citenp{press}) with this spectrum.  Of the
other 30 sources, only two had values of prob(\chisqrnu)$>$\ee{-6},
while the remainder were far below this value and are thus strongly
rejected.  We include the \chisqrnu\ values, and the best-fit
\kteffinfty\ values for sources \#3 and 31, in
Table~\ref{tab:sources}.

We then permit the value of \rinfty\ to vary; the results for these
fits are shown in Fig.~\ref{fig:qns}.  The majority (29 of 32) of
X-ray sources are spectrally hard, and have best-fit \rinfty $\ll$ 1
km. Only source \#3 overlaps with the (\kteffinfty,\rinfty) parameter
space of field qNSs. Two other spectrally soft objects (\#21 and \#31)
are observed, but do not produce \rinfty\ consistent with that of a
neutron star.  We show a pulse-height amplitude (PHA) spectrum for
source \#3 in Fig.~\ref{fig:obj3}.

 As shown previously (see references below), \chandra\ CCD
spectroscopy has confirmed that millisecond radio pulsars are
detectable in X-rays as power-law sources \cite{pooley00,grindlay01}.
We also modelled sources \#3 and 31 with an absorbed power-law. Source
\#3 did not produce an acceptable fit (\chisqrnu=12.8/2 dof;
prob=3\tee{-6}).  Source \#31 obtained an acceptable fit, but with a
photon power-law slope steeper than observed from MSPs (3.05\ppm0.35
vs. $\approx$1--2;
\citenp{saito97,zavlin98b,rots98,takahashi98,mineo00}).
(Alternatively, at least one MSP can be described as a thermal polar
cap with a H atmosphere spectrum, with a polar cap radius of $\sim$ 1
km [\citenp{zavlin98b}], considerably smaller than the H atmosphere
radii found from field sources).  We exclude an MSP origin for both
objects.  

To compare the spectra with those of CVs, we obtained best-fit
optically thin thermal bremsstrahlung (OTTB) spectral fits for sources
\#3 and 31 with \nh\ fixed at its optically implied value, extracting
the model parameters \ktbremss\ and $EM$(=$\int n_e n_p dV$) for
comparison with values of other CVs from Eracleous \etal\
(\citenp*{eracleous91}; E91 hereafter).  Results are shown in
Fig.~\ref{fig:cvs}.  Source \#3 is spectrally softer than all CVs
observed by E91, while having the second highest $EM$.  Source \#31 is
consistent with being inside the CV parameter space.  We exclude a CV
origin for source \#3, and note the spectral consistency of source
\#31 with a CV.

  For sources \#3 and \#31, we also fit 2 temperature Raymond-Smith
(2T RS) spectral models, as typically observed from RS~CVns
\cite{swank81,lemen89,dempsey93b}, holding $Z=0.019$\zsun, which is
the mode of metalicities from cluster stars \cite{majewski99,lee99}.
We compare our 2T RS spectral parameters ($kT_1$, $EM_1$, $EM_2/EM_1$)
with the spectral parameters for $\sim$40 field RS~CVns measured by
\acite[D93 hereafter]{dempsey93b}.  We hold the high temperature
$kT_2$ fixed at the average value found by D93 (1.4 keV), and permit
the remaining parameters to vary.  We find no acceptable parameter
space within (0.1$<kT_1<1.0$ keV, $EM_1<$\tee{54} \perval{cm}{-3}, and
$EM_2/EM_1>$1) in source \#3 (best fit values: $kT_1=0.42\pm0.03$ keV,
$EM_1$=1170\ud{190}{160} \ee{53} \perval{cm}{-3}; $EM_2/EM_1<0.016$
90\% confidence upper-limit).  In Fig.~\ref{fig:rscvns}, we show the
best-fit spectral parameters of sources \#3 and \#31 with those of
D93. These are clearly distinct from the spectral parameters of
RS~CVns, while in source \#31, they are perhaps comparable.

The NLSy1s are unusual as AGNs, as they have soft X-ray energy spectra
 \cite{boller96} .  The prevalence of NLSy1s in X-ray fields has been
 estimated at the $\sim$few per cent level \cite{hasinger00}, which is
 consistent with 1-2 such objects in the \ngc\ field.  NLSy1s are
 typically (although not always) acceptably modelled with a single
 photon power-law of slope $\alpha$=3-4, with no absorption above that
 due to the Milky Way, even of high S/N X-ray spectra observed with
 \rosat/PSPC, such as I~Zw~1 and Mrk~1044A \cite{boller96}.  Source
 \#3, however, is not acceptably fit with a pure power-law spectrum at
 the galactic absorption (see above), and therefore it would have an
 unusual X-ray spectrum for a NLSy1.  In addition, accepting the
 hypothesis that source \#3 is a NLSy1 requires accepting as a
 coincidence that the spectral intensity corresponds to an \rinfty=13
 km object for a H atmosphere spectrum.  Both the spectrum and
 intensity are explained naturally under the qNS hypothesis, and we
 therefore favor the interpretation of source \#3 as a qNS.  A
 definitive exclusion of a NLSy1 origin will require HST imaging and
 spectroscopy of objects in the X-ray error box.  Presently available
 HST observations of \ngc\ do not include source \#3.

The 3$\sigma$ upper-limit on the 0.5-10 keV luminosity of an
$\alpha=1$ power-law component is $<$12\% of the thermal component
luminosity in the same passband.  This is below observed values from
\cenx4\ and \aql \cite{asai96b,rutledge01,campana00,rutledge01b}.  The
absence of a stronger power-law component is due in part to our
spectral selection, requiring \rinfty=13 km thermal sources. 


\subsection{Astrometric Correction}
Carson \etal\ (\citenp*{carson00}; C00 hereafter) examined three
ROSAT/HRI sources (A, B and C) in the GC core, and optically
identified CVs spatially coincident with two of them. Their Star A
(Table 1 in C00) corresponds to our source \#4; their Star B is our
source \#6.  The optical separation of C00's CVs is
38.0\ppm0.2\arcsec, compared with the X-ray separation of the
\chandra\ sources of 37.99\ppm0.03\arcsec.  We therefore re-assign
astrometry on the basis of C00's positions.  This astrometry is
reflected in the source positions given in Table~\ref{tab:sources}.
Prior to re-assignment, the astrometry of star A and B were offset
from the \chandra\ pointing astrometry by (0.0,0.6\arcsec) and
(0.0,0.8\arcsec) in (R.A., dec.), consistent with the absolute
pointing accuracy of \chandra.  This results in a systematic
positional uncertainty of 0.15\arcsec\ for \source3, limited by the
astrometric accuracy of C00.

\section{Properties of the Source \#3} 

We have found in the previous section that the spectral and intensity
properties of source \#3 are readily explained if the source is a
qNS.  We examine the source characteristics in more detail in this
section. 

Its position at 1.7 core radii is perhaps surprising given the
expectation of mass segregation expected in the core of GCs
\cite{verbunt88}.  However, timing solutions for 15 millisecond
pulsars in 47 Tuc place eight of these beyond 1.7 core radii
($r_c$=23.1\ppm1.7\arcsec; \citenp{freire01}), indicating that such a
large radius is not uncommon for neutron stars in GCs, as may be
expected from exchange interactions with primordial cluster binaries
\cite{hut91}.

The X-ray sources in \ngc\ have been observed with {\em Einstein},
\rosat/PSPC and \rosat/HRI and examined several times
\cite{hertz83,johnston94,cool95a,verbunt00,verbunt01}. In some earlier
work our source \#3 is their source B \cite{hertz83,cool95a}.
Elsewhere, it is source \#7 \cite{johnston94,verbunt00,verbunt01}.
None of these works speculated on the classification of (our) source
\#3.  While an earlier paper found this source to be extended
\cite{cool95a}, later, more sensitive work found the object was a point
source \cite{verbunt00}, as do we.

\acite[V00 hereafter]{verbunt00} examined 7 HRI observations, taken at
7 different observational epochs (Aug 1992, Jan 1993, Jul 1994, Jan
1995, Jul 1995, Feb 1996 and Jan 1997), and found no evidence for
variability in source \#3.  Dividing the longest observation (74 ksec)
into two parts (separated by 10 days), V00 finds the countrates of
(1.7\ppm0.3 and 1.1\ppm0.2 c/ksec), which are different by a factor of
1.5\ppm0.4 (that is, consistent with being the same, within a factor
of $<$2.7, 3$\sigma$).  In comparison, a factor of \approxlt 3
variability over timescales of days \cite{campana97} to years
(\citenp{garcia99}; Rutledge \etal\ 2000, 2001a, 2001b) in field qNSs
has been observed. The absence of variability in \source3\ on a
$\sim$5 year timescale is consistent with the quiescent luminosity
being 100\% due to a hot thermal core of a qNS.

Finally, we extracted the X-ray  counts of source \#3 from the
120 ksec of publicly available HRI data
\footnote{obtained from http://heasarc.gsfc.nasa.gov}; the majority of
the integration was collected from three observations (1995 Jan, 17.7
ksec; 1995 Jul, 75.5 ksec; 1996 Feb, 13.0 ksec), collected $\approx$5
yrs prior to the \chandra\ observation.  A joint spectral fit between
the \chandra\ spectrum and the HRI observation, using an absorbed H
atmosphere spectrum, finds the average HRI countrate is consistent
with that predicted by the \chandra\ spectrum, with a ratio of HRI
countrate/ACIS predicted countrate of 1.14\ppm0.12 (1$\sigma$; or a
limit of $<$50\% variability at 3$\sigma$ confidence).

The number of counts per bin in all 21448 time bins (duration
$\approx$3.24 sec) is consistent with a Poisson distribution.  A
power-density spectrum \cite{rogg} of the joined observations also show
no power in excess of the Poisson level, with a 3$\sigma$ upper limit
of $<$25\% rms (0.2-10 keV; 10-\ee{5} sec). 
Thus, source \#3 exhibits no evidence of intensity variability on
timescales between 4 sec and 5 yr. 

\section{Discussion and Conclusions}

Source \#3 (\source3) is spectrally consistent with being a quiescent
neutron star, and not with being a CV, RS~CVn, or MSP.  The resulting
normalization in a H atmosphere spectral interpretation gives
\rinfty=14.3\ppm2.1 (D/ 5 kpc) km as a coincidence.  The uncertainty
in distance to \ngc\ is conservatively estimated at 10\%
(S. G. Djorgovski 2001, priv. comm.), which, when we include this in
our quoted uncertainty, gives a NS radius of 14.3\ppm2.5 km.  We give
its best fit H atmosphere spectral parameters in
Table~\ref{tab:values}, along with those from the field sources from
our previous studies.  The observed bolometric thermal luminosity is
$L_{\rm bol,\infty}=$(5\ppm2)\tee{32} \cgslum.  The low value of
\kteffinfty\ -- comparable to that of \cenx4\ -- may indicate the two
have similar outburst timescales ($>$30 yr in the case of \cenx4)
assuming similar outburst accretion rates, (cf. Eq 1) which may
explain the absence of a recorded outburst from \source3.

If source \#3 were a NLSy1 source, the X-ray spectrum is unusual, in
that it is not consistent with a single power-law slope, and we must
additionally accept as a coincidence that the intensity should be the
same as from a 10 km neutron star at the distance of \ngc.  Caution
must be taken, however, as the source was selected on the basis of its
consistency with this radius; it remains to be unequivocally
demonstrated that the object in question is indeed a qNS.

 We cannot explicitly exclude that source \#3 is an unusual MSP, CV,
RS~CVn or a NLSy1, particularly given that the number of X-ray sources
detected in \ngc\ are comparable to that expected from background
AGN. Nonetheless, we conclude the object is most likely a typical
example of a field qNSs, rather than an unusual example of another
population.

The classification of source \#3 as a qNS can be confirmed with an
optical identification of the companion, through ellipsoidal
variations associated with the binary orbital period, and by optical
spectroscopy revealing an accretion disk \cite{mcclintock00}.  Naturally,
an X-ray outburst would also confirm this classification. If
further observations of \source3\ support its identification as a qNS,
this would then be the first qNS identified in quiescence through its
spectral properties, rather than through an X-ray outburst. 

The apparent absence of a power-law spectral component which has been
observed from field qNSs \cenx4\ and \aql\ may indicate that this
spectral component is not always present in qNSs.  It has been
suggested that the observed variability in field qNSs is due
exclusively to the power-law spectral component \cite{rutledge01}.  If
so, then the absence of both the power-law component and intensity
variability in source \#3 is consistent with its identification as a
qNS.  Because we do not include a power-law component in our spectral
modeling, we may have missed additional qNSs in this GC, which are
nonetheless detected with \chandra.

 The identification of previously unknown qNSs in globular clusters
through their quiescent X-ray spectra opens up new observational
opportunities for this class of sources.  As discussed in
\acite{rutledge00}, the distances to GCs post-{\em Hipparcos} can be
measured to as little as 3\% uncertainty (see, for example,
\citenp{carretta00}).  This is far superior to the sometimes 50-100\%
uncertainty in the source distances for field qNSs.  Combined with an
accurate measurement of the qNS angular size obtained with X-ray
spectroscopy of the thermal H atmosphere, one then can produce
precision measurements of NS radii.  The value of such measurements
was highlighted recently by \acite{lattimer01}, who showed that one NS
radius measurement of \ppm1 km accuracy could potentially exclude
nearly half of the proposed nuclear equations of state.

The radius measurements of NSs in GCs would be limited in accuracy by
the signal-to-noise of the spectrum and the systematic uncertainty in
the intrinsic spectrum, rather than the uncertainty in source
distance.  The backside-illuminated ACIS-S detector on \chandra\
(which has higher sensitivity at energies below 1 keV) would produce a
factor of 2.5 higher countrate than the ACIS-I for the observed
spectrum of \source3; in 70 ksec, this would measure \rinfty/D to
8\%(1$\sigma$) if \nh\ is assumed {\em a priori} (\ppm20-35\% if one
does not assume the \nh\ value).  The {\em XMM}/pn detector would
produce a factor of 5 higher countrate than ACIS-I, over a similar
passband, with subsequent improvement on the derived \rinfty/D.  With
{\em Con-X}, \source3\ would produce 11000 counts in a 30 ksec
observation sufficient to measure \rinfty/D to \ppm7\%, including
uncertainty in all spectral parameters (\nh, \kteff, \rinfty/D, and a
power-law $\alpha$ and normalization).  Combined, with a 3\%
uncertainty in distance to the GC, \rinfty\ would be measured to
$\ppm$ 8\% (\ppm1 km for \rinfty=13 km).  Thus, a program measuring
the radii of transient neutron stars in quiescence in GCs would
usefully constrain the NS equation of state. 

\acknowledgements

The authors continue to be grateful to the \chandra\ Observatory team
for producing this exquisite observatory. This research was partially
supported by the National Science Foundation under Grant
No. PHY99-07949 and by NASA through grant NAG 5-8658 and NAG 5-7017.
GGP acknowledges support through NAG5-10865. 
L. B. is a Cottrell Scholar of the Research Corporation.
E. F. B. acknowledges support from an Enrico Fermi Fellowship.

\newpage

\newpage

\pagestyle{empty}
\begin{figure}[htb]
\caption{\label{fig:obj3} The $\nu F_\nu$ model spectrum of
\source3\ (Source \#3), observed with the ACIS-I.  The solid line is
the best-fit unabsorbed H atmosphere model spectrum, with \nhtt=0.09
held fixed (that is, the intrinsic X-ray spectrum, prior to absorption
by the inter-stellar medium; see Table~\ref{tab:values}).  The cross
points are the background subtracted observational PHA spectrum,
covering the energy ranges, with the error bars taken from count
rates.   }
\end{figure}

\pagestyle{empty}
\begin{figure}[htb]
\caption{\label{fig:qns} Best-fit H atmosphere spectral parameters for
spectroscopy sources, compared with the field qNSs. Points are marked
with source number, and 1$\sigma$ error bars (field sources have 90\%
confidence error bars) are the best-fit \kteffinfty\ and \rinfty, with
\kteffinfty\ limited for this fit at 410 eV.  The broken line marks
the approximate flux limit for a H atmosphere source ($\propto$
\rinfty$^2$\kteffinfty$^4$).  Source \#3 overlaps in \rinfty\ with field
qNSs \aql, \cenx4, 2129+47 and 1608-522.  Source numbers in the
upper-right box did not produce statistically acceptable fits. The
connected points for 2129+47 and 1608-522 are for two different
assumptions of distance and \nh\ (see Rutledge \etal\ 1999, 2000).}
\end{figure}

\pagestyle{empty}
\begin{figure}[htb]
\caption{\label{fig:cvs} Comparison between the $EM$ (in arbitrary
units) vs. $kT_{\rm Bremss.}$ of field CVs from E91 (solid squares),
and source \#3 and \#31 (as marked).  Source \#3 is spectrally softer
than all other CVs observed by E91, and has the highest EM. }
\end{figure}

\pagestyle{empty}
\begin{figure}[htb]
\caption{\label{fig:rscvns} Comparison of $EM_1$ vs. $EM_2/EM_1$
between RS~CVns (D93) and sources \#3 and \#31.  The RS~CVn $EM_2/EM_1$
values are the best-fit values (D93), while the qNSs' $EM_2/EM_1$
values are 90\% confidence upper-limits.  The implied values of $EM_1$
for all but \#36 are significantly higher than observed from field
RS~CVns. The upper-limits $EM_2/EM_1$ are below the values measured
from field RS~CVns in source \#3, but not source \#31.  }
\end{figure}

\clearpage
\pagestyle{empty}
\begin{figure}[htb]
\PSbox{fig1.ps hoffset=-80 voffset=-80}{14.7cm}{21.5cm}
\FigNum{\ref{fig:obj3}}
\end{figure}

\clearpage
\pagestyle{empty}
\begin{figure}[htb]
\PSbox{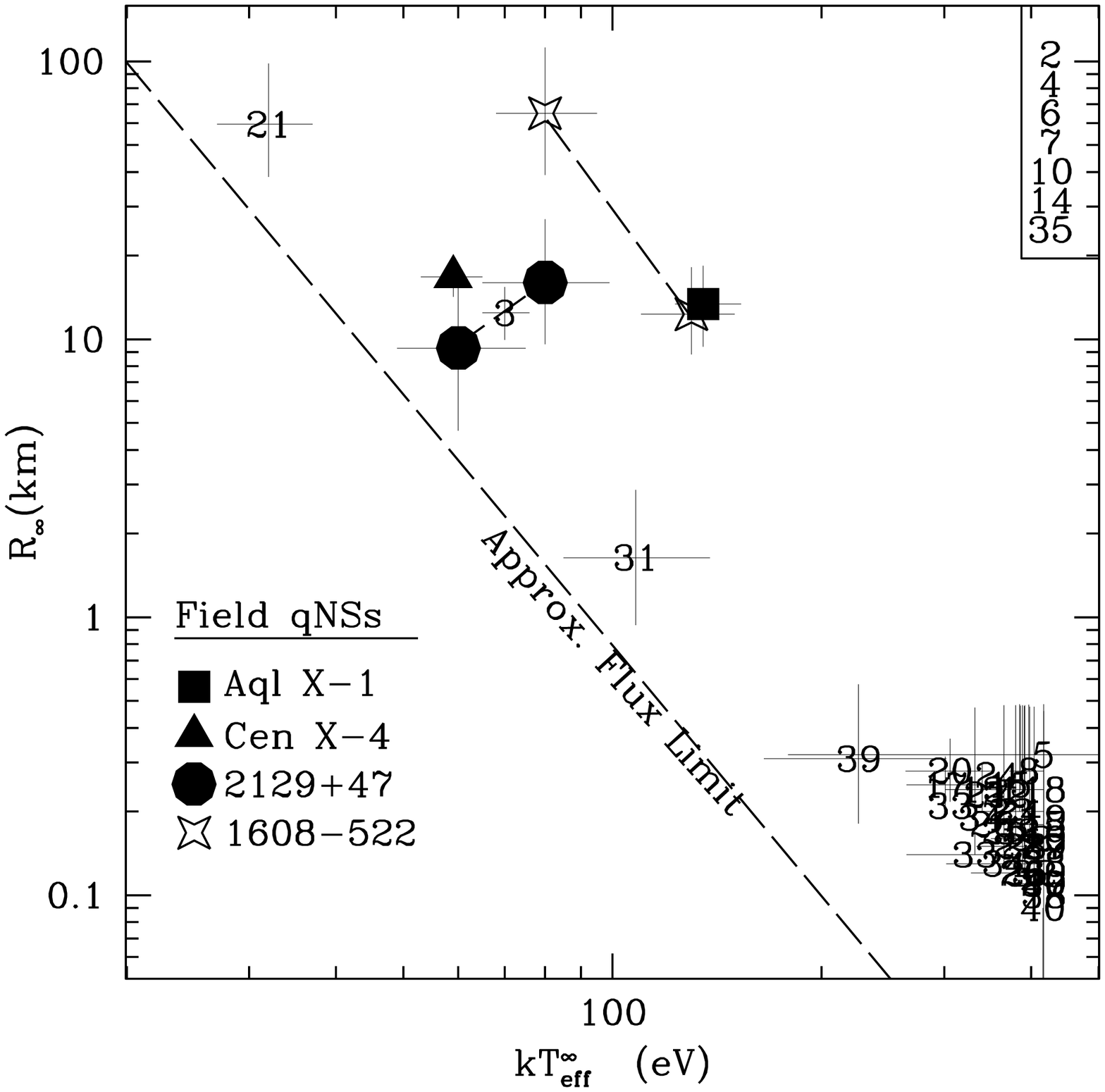 hoffset=-80 voffset=-80}{14.7cm}{21.5cm}
\FigNum{\ref{fig:qns}}
\end{figure}

\clearpage
\pagestyle{empty}
\begin{figure}[htb]
\PSbox{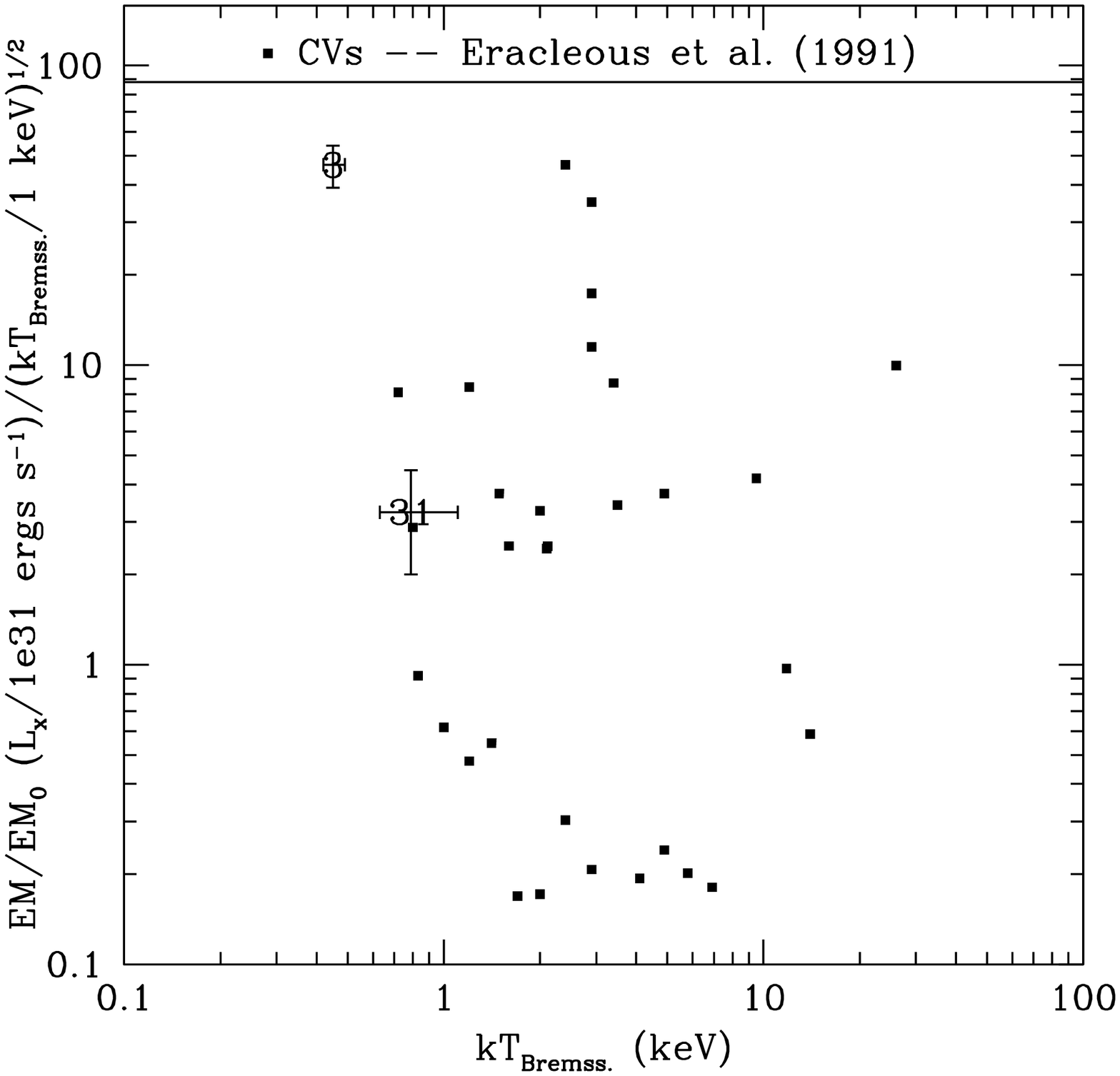 hoffset=-80 voffset=-80}{14.7cm}{21.5cm}
\FigNum{\ref{fig:cvs}}
\end{figure}

\clearpage
\pagestyle{empty}
\begin{figure}[htb]
\PSbox{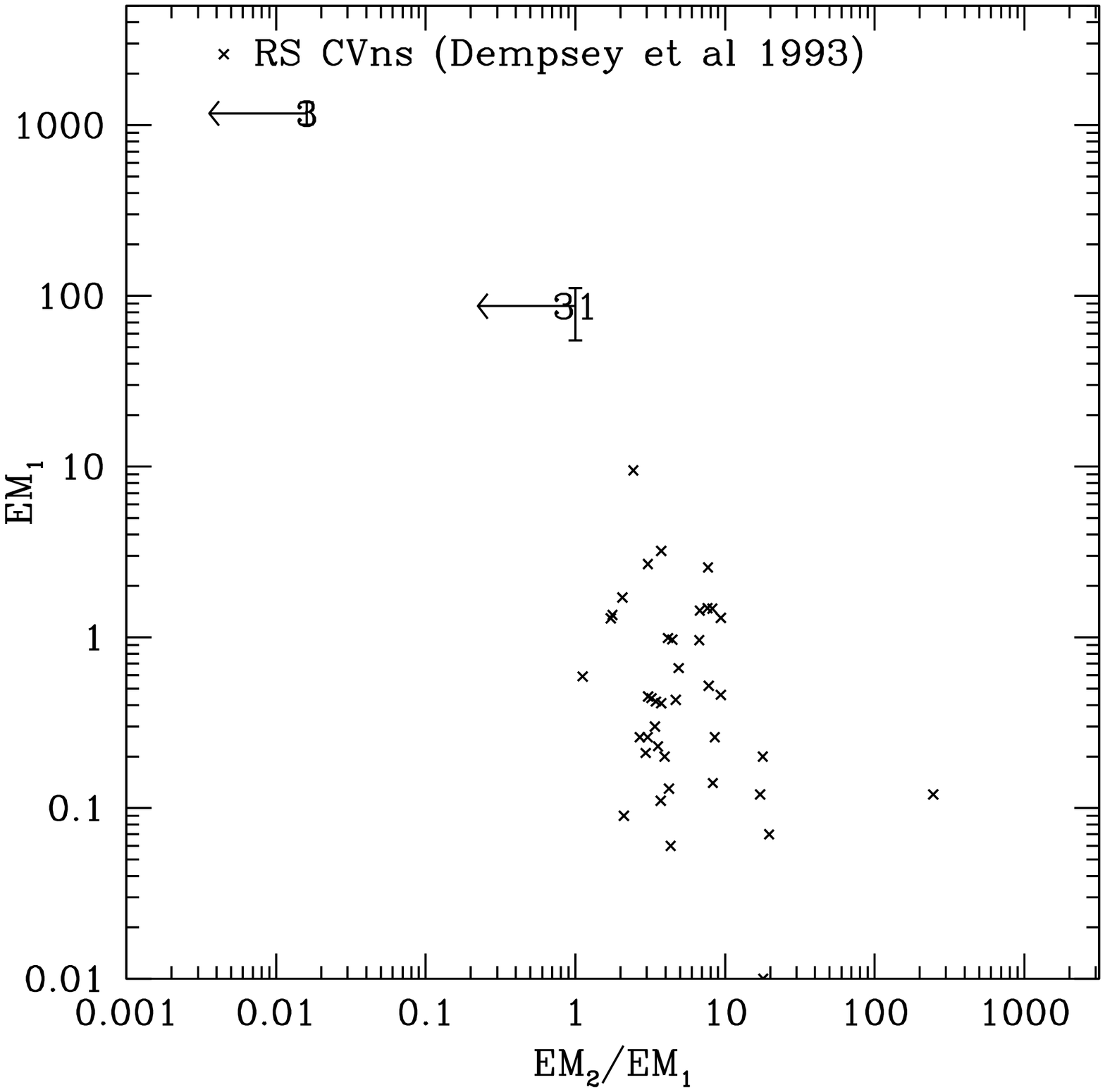 hoffset=-80 voffset=-80}{14.7cm}{21.5cm}
\FigNum{\ref{fig:rscvns}}
\end{figure}

\SingleSpace

\begin{deluxetable}{lrrrrrrlrcl}
\scriptsize
\tablecaption{\label{tab:sources} \chandra\ X-ray Sources in the Field of NGC 5139}
\tablewidth{17cm}
\tablehead{
\colhead{} & \colhead{} & \colhead{R.A. } & \colhead{Dec. } & \colhead{Stat. $\delta^a$} &\colhead{} &\colhead{} &\colhead{{\kteffinfty}} &\colhead{I} &\colhead{} &\colhead{} \\
\colhead{ID} & \colhead{SNR} & \colhead{(J2000)} & \colhead{(J2000)} & \colhead{(\arcsec)} & \colhead{$\Delta/r_c$~$^e$} & \colhead{\chisqrnu/dof (prob)} &\colhead{(eV) } & \colhead{c/ksec (\ppm)}&\colhead{Type$^b$}   & \colhead{ref.} 
}
\startdata
  1 & 53.3 & 13 25 52.211 & -47 19  8.05 &           \ppm0.17   &  5.04 &   \nodata$^d$	             &                -- & 50.7 (1.6)    & dMe& 1     \\ 
  2 & 21.6 & 13 26  1.531 & -47 33  6.67 &    \ppm0.14/\ppm0.18 &  3.35 &   226/3 (n/a)              &                -- & 13.4 (0.5)    &  --&     \\ 
  3 & 20.1 & 13 26 19.757 & -47 29 11.51 &    \ppm0.03/\ppm0.05 &  1.71 &  2.61/3 (0.050)            &        72\ppm    1&  6.9 (0.4)    & qNS& 3    \\ 
  4 & 19.8 & 13 26 52.134 & -47 29 36.29 &             \ppm0.02 &  0.56 &   105/3 (n/a)              &                -- &  9.9 (0.4)    &  CV& 2    \\ 
  5 & 18.4 & 13 26 54.496 & -47 22  5.34 &    \ppm0.10/\ppm0.08 &  2.57 &    78/3 (n/a)              &                -- &  7.9 (0.4)    &  --&     \\ 
  6 & 17.5 & 13 26 53.503 & -47 29  0.92 &             \ppm0.02 &  0.52 &   117/3 (n/a)              &                -- & 10.2 (0.4)    & CV &2   \\ 
  7 & 16.2 & 13 27 29.326 & -47 25 54.92 &    \ppm0.16/\ppm0.12 &  3.00 &    44/3 (9\tee{-29})       &                -- &  6.0 (0.3)    &  --&     \\ 
  8 & 15.0 & 13 27 12.887 & -47 34 57.79 &             \ppm0.17 &  3.00 &    51/3 (\ee{-33})         &                -- &  5.3 (0.3)    &  --&     \\ 
$^c$  9 & 14.5 & 13 25 30 & -47 15 24 &                 \nodata &  7.05 &  \nodata$^d$               &              --   &  7.3 (0.7)    &  --&     \\ 
 10 & 13.3 & 13 26 20.336 & -47 30  3.92 &    \ppm0.06/\ppm0.08 &  1.75 &    76/3 (n/a)              &                -- &  4.5 (0.3)    &  --&     \\ 
 11 & 12.8 & 13 27 11.785 & -47 32 41.43 &    \ppm0.12/\ppm0.10 &  2.30 &    41/3 (\ee{-26})         &                -- &  3.8 (0.3)    &  --&     \\ 
 12 & 11.7 & 13 26 11.503 & -47 34  4.81 &             \ppm0.28 &  3.06 &   \nodata$^d$              &                -- &  2.9 (0.2)    &  --&     \\ 
 13 & 10.3 & 13 26 44.106 & -47 32 32.31 &             \ppm0.07 &  1.51 &    24/3 (\ee{-15})         &                -- &  2.5 (0.2)    &  --&     \\ 
 14 & 10.1 & 13 27 27.495 & -47 31 33.41 &             \ppm0.21 &  2.92 &  8.57/3 (\ee{-05})         &                -- &  2.7 (0.2)    &  --&     \\ 
 15 &  9.6 & 13 26 25.089 & -47 32 28.51 &    \ppm0.09/\ppm0.12 &  2.00 &    24/3 (4\tee{-16})       &                -- &  2.5 (0.2)    &  --&     \\ 
$^c$ 16 &  9.3 & 13 25 38.4 & -47 17 51 &             \nodata   &  6.03 &  \nodata$^d$               &               --  &  4.3 (0.5)    &  --&     \\ 
 17 &  9.0 & 13 26 37.412 & -47 30 53.95 &             \ppm0.05 &  1.04 &    20/3 (4\tee{-13})       &                -- &  1.7 (0.2)    &  --&     \\ 
 18 &  8.7 & 13 26 48.640 & -47 27 45.75 &             \ppm0.03 &  0.37 &   \nodata$^d$              &                -- &  2.9 (0.2)    &  --&     \\ 
 19 &  8.5 & 13 26 41.448 & -47 22 16.79 &             \ppm0.18 &  2.45 &    33/3 (2\tee{-21})       &                -- &  2.0 (0.2)    &  --&     \\ 
 20 &  8.4 & 13 25 57.393 & -47 32 51.72 &             \ppm0.47 &  3.54 & 14.78/3 (\ee{-09})         &                -- &  2.0 (0.2)    &  --&     \\ 
 21 &  8.1 & 13 26 46.014 & -47 19 45.40 &    \ppm0.52/\ppm0.40 &  3.40 &  7.57/3 (5\tee{-05})       &                -- &  1.8 (0.2)    &  --&     \\ 
 22 &  7.8 & 13 26 27.510 & -47 34 57.66 &             \ppm0.29 &  2.71 &    22/3 (8\tee{-15})       &                -- &  1.9 (0.2)    &  --&     \\ 
 23 &  7.6 & 13 26 48.717 & -47 31 26.09 &             \ppm0.07 &  1.10 &    18/3 (3\tee{-12})       &                -- &  1.6 (0.2)    &  --&     \\ 
 24 &  7.4 & 13 26 23.500 & -47 19 22.09 &    \ppm0.48/\ppm0.66 &  3.84 &    18/3 (5\tee{-12})       &                -- &  2.3 (0.2)    &  --&     \\ 
 25 &  6.2 & 13 26 34.310 & -47 30 34.50 &             \ppm0.08 &  1.06 & 11.76/3 (\ee{-07})         &                -- &  1.0 (0.1)    &  --&     \\ 
 26 &  6.2 & 13 26 55.078 & -47 31 14.49 &             \ppm0.09 &  1.17 & 14.89/3 (\ee{-09})         &                -- &  1.0 (0.1)    &  --&     \\ 
 27 &  6.2 & 13 27 21.200 & -47 23 23.73 &             \ppm0.38 &  3.05 &    18/3 (6\tee{-12})       &                -- &  1.2 (0.2)    &  --&     \\ 
 28 &  6.1 & 13 26 13.561 & -47 34 40.63 &             \ppm0.45 &  3.13 &      \nodata$^d$           &                -- &  1.2 (0.2)    &  --&     \\ 
 29 &  6.1 & 13 26 51.057 & -47 31 45.64 &             \ppm0.08 &  1.25 &    16/3 (5\tee{-11})       &                -- &  1.1 (0.1)    &  --&     \\ 
 30 &  6.0 & 13 26 39.240 & -47 36 30.25 &             \ppm0.65 &  3.06 &    \nodata$^d$             &                -- &  1.6 (0.2)    &  --&     \\ 
 31 &  6.0 & 13 26 38.258 & -47 19 57.00 &             \ppm0.65 &  3.36 &  1.70/3 (0.165)            &        50\ppm1    &  1.0 (0.2)    &  --&      \\ 
 32 &  5.8 & 13 27 28.290 & -47 24 24.86 &    \ppm0.55/\ppm0.44 &  3.19 & 11.78/3 (\ee{-07})         &                -- &  1.0 (0.1)    &  --&     \\ 
 33 &  5.5 & 13 26 49.559 & -47 32 13.62 &             \ppm0.12 &  1.41 &  8.65/3 (\ee{-05})         &                -- &  0.7 (0.1)    &  --&     \\ 
 34 &  5.5 & 13 27  7.998 & -47 23 34.23 &             \ppm0.22 &  2.41 & 11.77/3 (\ee{-07})         &                -- &  0.8 (0.1)    &  --&     \\ 
 35 &  5.5 & 13 27 10.001 & -47 33 21.99 &    \ppm0.26/\ppm0.36 &  2.40 &    23/3 (3\tee{-15})       &                -- &  1.4 (0.2)    &  --&     \\ 
$^c$ 36 &  5.5 & 13 25 42.7   & -47 19 16 &        \nodata   	 &  5.45 &  \nodata$^d$              &                -- &  1.3 (0.3)    & -- &     \\ 
 37 &  5.4 & 13 26 26.964 & -47 34 10.57 &    \ppm0.24/\ppm0.33 &  2.46 &    17/3 (4\tee{-11})       &                -- &  1.1 (0.1)    &  --&     \\ 
 38 &  5.1 & 13 26 59.191 & -47 34 59.26 &             \ppm0.36 &  2.60 &  9.04/3 (6\tee{-06})       &                -- &  1.0 (0.1)    &  --&     \\ 
 39 &  5.1 & 13 26 12.686 & -47 24 13.71 &    \ppm0.31/\ppm0.41 &  2.74 &  6.00/3 (4\tee{-04})       &                -- &  0.9 (0.1)    &  --&     \\ 
 40 &  5.1 & 13 27  6.416 & -47 25 38.76 &    \ppm0.22/\ppm0.18 &  1.75 &  9.00/3 (6\tee{-06})       &                -- &  0.7 (0.1)     &  --&     \\ 
\enddata 
\tablecomments{
$^a$~Statistical uncertainty in R.A./Dec. (or, both), in arcsec. Systematic positional uncertainty is \ppm0.15\arcsec\ (see text). 
$^b$~Type: qNS= quiescent transient neutron star; CV= cataclysmic variable; 
$^c$~X-ray {\em celldetect} positional uncertainty possibly incorrect by up to 10\arcsec due to large off-axis angle
and low S/N; 
$^d$ Object not spectrally analysed (see text); 
$^e~r_c$=156\arcsec \cite{djorgovski93}; 
1, \acite{cool95a};
2, \acite{carson00}; 
3, present work; 
}
\end{deluxetable}

\begin{deluxetable}{lrllll}
\tablecaption{\label{tab:values} H Atm. Spectral Parameters for \source3\  and Field qNSs }
\tablehead{
\colhead{}	& \multicolumn{2}{c}{\rinfty}	&
\colhead{\kteffinfty}	&\colhead{\nh}	& \colhead{} \\
\colhead{Object}	&\multicolumn{2}{c}{(km)}	&
\colhead{(eV)}	&\colhead{(\ee{22} \perval{cm}{-2})}	& \colhead{Ref.} 
}
\startdata
\source3\ & 14.3\ppm2.1 & (D/5 kpc) & 66\ud{4}{5}& (0.09)& 0\\
\aql\ 	& 13.4\ud{5}{4} &(D/5 kpc) & 135\ud{18}{12} &
0.35\ud{0.08}{0.07}& 1\\
\cenx4\	& 16.8\ppm2.6 &(D/1.2 kpc)    & 59\ppm6 & (0.055)
& 2\\
$^a$ 4U 2129+47& 9.3\ud{7.8}{4.6} &(D/1.5 kpc)& 60\ud{15}{11} & (0.28) & 3\\
$^a$ 4U 2129+47& 16\ud{11}{6.4} &(D/6.0 kpc)& 80\ud{19}{15} & (0.17) & 3\\
$^a$ 4U 1608-522 & 12.3\ud{5.9}{3.5} &(D/3.6 kpc)& 130\ppm20	& (0.8) & 4 \\
$^a$ 4U 1608-522 & 65\ud{47}{26} &(D/3.6 kpc)& 80\ud{15}{12}	& (1.5) & 4 \\
\enddata
\tablecomments{
Error bars are 90\%
confidence.  Values in parenthesis are assumed.  Previously measured
values have been converted from $R$ and $kT_{\rm eff}$ to \rinfty\ and
\kteffinfty\ (values as appear at infinity); 
0, present work; 
1, \acite{rutledge01b}; 
2, \acite{rutledge01}; 
3, \acite{rutledge00}; 
4, \acite{rutledge99}; 
$^a$ Different assumed values of D and/or \nh. }
\end{deluxetable}

\end{document}